# Microwave-free J-driven DNP (MF-JDNP): A proposal for enhancing the sensitivity of solution-state NMR

Maria Grazia Concilio[*], and Lucio Frydman[*]

*Department of Chemical and Biological Physics, Weizmann Institute of Science, Rehovot, Israel*

J-driven Dynamic Nuclear Polarization (JDNP) was recently proposed for enhancing the sensitivity of solution-state nuclear magnetic resonance (NMR), while bypassing the limitations faced by conventional (Overhauser) DNP at magnetic fields of interest in analytical applications. Like Overhauser DNP, JDNP also requires saturating the electronic polarization using high-frequency microwaves –known to have poor penetration and associated heating effects in most liquids. The present microwave-free JDNP (MF-JDNP) proposal seeks to enhance solution NMR's sensitivity by shuttling the sample between higher and lower magnetic fields, with one of these fields providing an electron Larmor frequency that matches the inter-electron exchange coupling $J_{ex}$. If spins cross this so-called JDNP condition sufficiently fast, we predict that a sizable nuclear polarization will be created without microwave irradiation. This MF-JDNP proposal requires radicals whose singlet/triplet self-relaxation rates are dominated by dipolar hyperfine relaxation, and shuttling times that can compete with these electron relaxation processes. This communication discusses the theory behind the MF-JDNP, as well as proposals for radicals and conditions that could enable this new approach to NMR sensitivity enhancement.

*Emails: maria-grazia.concilio@weizmann.ac.il, lucio.frydman@weizmann.ac.il

## I. Introduction

Overhauser Dynamic Nuclear Polarization (ODNP) can enhance the sensitivity of solution state NMR by saturating an electron radical comixed with the sample of interest.[1-3] However, unless aided by the contact couplings that occasionally arise in the presence considerable electron delocalization on the target nucleus,[4-10] ODNP is only efficient at low magnetic fields.[11-16] High-field ODNP experiments based intermolecular contact couplings have thus been reported on nuclei like $^{31}$P, [17, 18] $^{19}$F [4, 19] and $^{13}$C, [8, 20, 21] However, in the most general and analytically-relevant cases such as those involving $^{1}$Hs, the electron and nuclei will interact solely through intermolecular dipolar coupling. In a system in which a radical interacts with the solvent only through dipolar coupling, the DNP efficiency decays rapidly with magnetic field $B_0$. Indeed, typical $^{1}$H ODNP enhancements drop from a maximum of ≈330x when $B_o$ ≤ 0.4T, to ≈1.001x at the ≥7 T fields were contemporary NMR is done.[20, 22-25] The decreased efficiency of ODNP with magnetic field deprives solution NMR from the benefits that DNP has brought to solid state analyses.[26-29]

We have recently discussed a possible way to bypass these solution-state limitations, based on what we denominate the J-driven DNP (JDNP) effect.[30] JDNP requires stable biradicals with identical monomers and an inter-electron exchange coupling $J_{ex}$ close to the electron Larmor frequency $\omega_E$. As the JDNP condition $J_{ex} \approx \pm \omega_E$ is fulfilled, a difference between the relaxation rates for the two-electron singlet and triplet states which are dipolar hyperfine-coupled to nuclear α or β states can lead, upon electron irradiation, to a transient imbalance between these nuclear populations. This in turn leads to nuclear magnetization enhancement. The physics of the JDNP is reminiscent of that observed in chemically induced dynamic nuclear polarization (CIDNP)[31-36] –an experiment in which a laser or a chemical reaction will drive the system away from the thermal equilibrium. The main JDNP requirements are thus stable biradicals, an inter-electron $J_{ex}$ in the order of $\omega_E$, and an efficient microwave irradiation at the electron Larmor frequency. Many radicals groups could serve as the starting point for the synthesis of such biradicals whose $J_{ex}$ would reach 100s GHz.[37-39] Exchange couplings in the 90-290 GHz range, for instance, have been obtained by linking bistrityl-based radicals. [39]

Electron saturation at such frequencies, however, is problematic in terms of microwave availability, sample heating, microwave penetration –and even the fact that exact $J_{ex}$ values are hard to predict or measure in solutions. The present study discusses a shuttling-based proposal [40-44] that might bypass these limitations. The ensuing microwave-free JDNP (MF-JDNP) approach proposes to polarize the nuclear spins by shuttling the sample between a lower and higher magnetic field. Sample shuttling technologies have been used previously in DNP to enhance $^{13}$C signals in experiments involving optical pumped NV-centres in diamonds, as well as to increase the $^{1}$H and $^{13}$C high-field polarization after executing ODNP at low magnetic fields.[5, 41, 45] In the case of MF-JDNP we show that if either the starting or the final magnetic field in a two-field shuttling experiment fulfills the $J_{ex} \approx \pm \omega_E$ condition, nuclear polarization will be created.

## II. Spin systems and methodology

This study's calculations were performed using the Spinach software package [46] based on laboratory frame Hamiltonians, as no rotating-frame approximation with respect to the microwaves is *a priori* justified. The simulation code is provided in the Supporting Information. For simplicity the electron *g*-tensors were assumed identical, axially symmetric and colinear, as would result from radicals joined by a short, rigid linker; still, as in liquids these *g*-tensors could become non-colinear due to rotational and vibrational modes, the effect of *g*-tensor orientation on the enhancement is discussed in the Supporting Information. The electrons' relaxation and spin dynamics were described using the singlet and triplet basis sets suitable for this $\Delta\omega \ll J_{ex}$ scenario. Spin population operators corresponding to the $\alpha$ and $\beta$ nuclear components of these singlet and the triplet states were considered, and described using Vega's fictitious operators notation.[30, 47, 48] Three- and four-spin systems were considered, encompassing in all cases the aforementioned two-electron biradical plus protons. One of the protons was always assigned to a fluid medium, that would dynamically diffuse around the biradical as described below; this is the "solvent" $^1$H whose polarization enhancement MF-JDNP is seeking, and which was assumed to interact with the electrons solely through dipolar (*aka* anisotropic hyperfine) couplings. Disregarding Fermi contact couplings is here justified by the fact that such protons would be located ≥5 Å away from the biradical's main electron density. A second proton with spatial coordinates fixed vs the electrons was occasionally included; the purpose of adding this "radical" $^1$H was to evaluate the detrimental effect that a proton belonging to the biradical, will have on MF-JDNP's ability to polarize the medium. Further details about the assumed systems are given in Table 1

To investigate the physics behind the MF-JDNP, brute-force numerical simulations accounting for every self- and cross-relaxation term within the Bloch-Redfield-Wangsness (BRW) relaxation theory, [49-51] were implemented in the Spinach software.[46] Considering that the exchange coupling has the same order of magnitude as the electron Larmor frequency, these calculations incorporated into the relaxation superoperator a scalar relaxation of the first kind.[52] According to Redfield's relaxation theory, and for the conditions mentioned above, the singlet and triplet relaxation rates of a biradical with identical *g*-tensors will be dominated by the dipolar hyperfine interactions between the electrons and the surrounding protons.[30] The lifetimes (T$_1$) for the $\hat{T}_{\pm,\alpha/\beta}$ states will then vary strongly with the distance between the electrons and the protons: when $^1$Hs do not approach the biradical electrons to distance closer than 5 Å, these T$_1$s extend into the ms range; [30] in the presence of "radical" protons sited ≤5 Å away from either of the electrons, these T$_1$s drop to ≈100 μs (see Supporting Information for the relaxation rates predicted by Redfield theory as a function of these and other parameters). This in turn posed the issue of how to estimate the relaxation behavior expected from two-electrons interacting with "solvent" protons, that can take a number of distances from the electrons. Figure 1 presents the model used to reproduce the expected behavior. Overall, we found that three regions can be distinguished for the behavior of this proton (vide infra). There is a "polarizing" region active when electron-nuclear distances are ca. 5-10 Å (labeled "A" in Fig. 1A), where the JDNP is active and also relaxation times shorten. Then there is an "outer" region happening when the $^1$H is ≥10Å away from the closest electron (labeled "B" in Fig. 1A), in which the nucleus will not undergo polarization effects, and electrons will only contribute to speed up the nuclear relaxation. This "B" region extends until the proton falls under the influence of another biradical, which for prototypical concentrations (<10 mM) will be sited some 30 Å away from the original biradical. Finally, there is a "close contact" region (labeled "C" in Fig. 1A) in which protons from the biradical itself reside, and which solvent protons will not be able to penetrate. In the case of a solvent proton, the nuclear spin will be diffusing randomly at ≈1 μm$^2$/ms (*i.e.* ≈10$^{-9}$ m$^2$ s$^{-1}$ diffusivity constant at 25 °C [11]), crossing several times in- and-out regions in which JDNP is active and regions where it is not. To account for this in our calculations, a periodic box of size 30x30x30 Å$^3$ centered on a biradical was therefore established, and a set of ten random walks in this 3D space were executed while counting how often the solvent proton diffuses in and out between the polarizing ("A") and non-polarizing ("B") regions to which it was allowed. Figure 1B shows an example of such walks, which reveal that the mean occurrence of a nuclear spin is about 10% of its time within the polarizing region "A", and the remaining 90% of its time in the "outer", non-polarizing volume, region "B". To account for this constant interruption of the polarization process in simulations, we set up a randomized polarizing scheme where proton coordinates were constantly exchanged between regions "A" and "B", with time steps of $1 \times 10^{-6}$ s and $9 \times 10^{-6}$ s respectively. Within each of these regions the total electron/nuclear spin ensemble was then allowed to relax according to Redfield's relaxation superoperator theory [50, 51, 53, 54]; for the sake of simplicity the nuclear coordinates were chosen fixed in this exchange model, with values representative of what numerous random walks simulations yielded as average of the full volumes of the "A" and "B" configurations. Table 1 presents these prototypical proton and electron Cartesian coordinates, as well as additional parameters used to create the relevant spin Hamiltonian. Notice that no electron relaxation arising from inter-radical dipolar interactions are here considered; if present, these could be prevented by lowering the polarizing agent concentrations [11, 55].

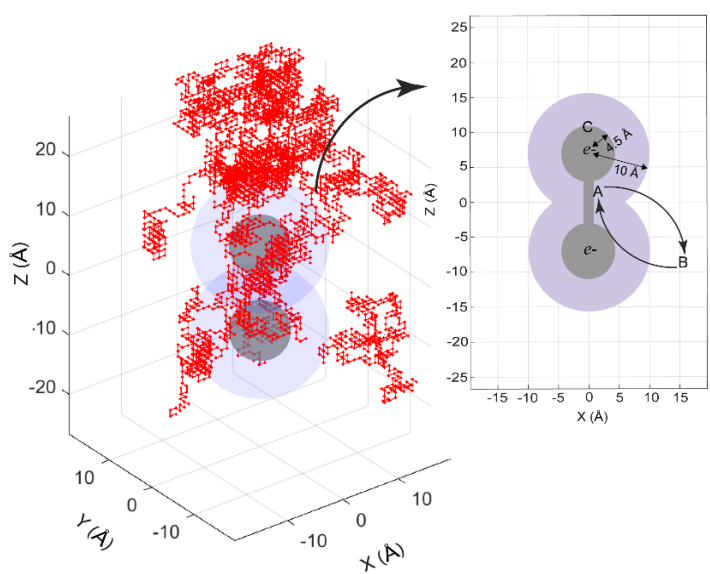

**FIG. 1:** Example of random walks undertook by a nuclear spin in a 3D box with size equal to 30x30x30 Å$^3$, with a diffusion constant of 1 μm/ms, corresponding to a random walk of 10000 steps of 1 Å in 1 ms. Periodic boundary conditions were set to represent an infinite system in which the nuclear spin can get out from a side of the box and get in from the other. In the insert: Polarization region accessible by the solvent (violet region) surrounding the biradical (grey circles connected by a linker); A and B represent two protons' configurations in and outside the polarization region, respectively, while C represents an intra-radical proton, the particles' coordinates were given in the Table 1. The code used to perform this random walk is provided in the supporting information.

**Table 1:** Biradical / protons spin system parameters used in the simulations.

| Parameter | Spin system |
|---|---|
| $^1$H chemical shift tensor eigenvalues, [xx yy zz], ppm | [5 5 5] |
| $^1$H chemical shift tensor, ZYZ active Euler angles, rad | [0.0 0.0 0.0] |
| Electron 1 *g*-tensor eigenvalues, [xx yy zz], Bohr magneton | [2.0032 2.0032 2.0026] |
| Electron 1 *g*-tensor, ZYZ active Euler angles, rad | [0.0 0.0 0.0] |
| Electron 2 *g*-tensor eigenvalues, [xx yy zz], Bohr magneton | [2.0032 2.0032 2.0026] |
| Electron 2 *g*-tensor, ZYZ active Euler angles, rad | [0.0 0.0 0.0] |
| $^1$H coordinates [x y z] / Å | "Solvent" $^1$H in region "A": [1.27,1.61,2.26]<br>"Solvent" $^1$H in region "B": [15.46,10.05,-8.0]<br>Radical proton in region "C": [0.0 0.0 10.4] |
| Electron 1 and electron 2 coordinates, [x y z] / Å | [0 0 -7.20] and [0 0 7.20], |
| Rotational correlation time $\tau_C$ / ns | 2.2 |
| Scalar relaxation modulation depth / GHz | 1 |
| Scalar relaxation modulation time / ps | 1 |
| Temperature / K | 298 |

## III. The MF-JDNP Effect

While ODNP/JDNP propose to polarize nuclei by taking electron spins out of equilibrium via microwave irradiation, MF-JDNP attempts to polarize nuclei without microwaves. Instead, it exploits the temporary imbalance that will occur in the electron polarization, if samples are suddenly moved along the axis of a finite solenoid magnet. We hypothesize that if such non-equilibrated electronic spins encounter the JDNP condition, the resulting relaxation process will lead to an imbalance between the $\alpha$ and $\beta$ nuclear components of the singlet and the triplet state – and in turn to NMR hyperpolarization. To explore this possibility numerous scenarios were envisioned; for simplicity we consider solely the one schematized in Fig. 2, where the sample is repeatedly shuttled between a high field where NMR measurements will be taken – for instance 14.1T, corresponding to a 600 MHz $^1$H Larmor frequency – and a field where the $J_{ex}\approx\pm\omega_E$ condition is fulfilled – for instance 9.1 T, corresponding to a $\omega_E \approx J_{ex} \approx$ -255 GHz.[37-39] Considering that contemporary pneumatic shuttling set ups can displace small samples with velocities of $\approx$40 m/s,[56] and that the distance between the two positions in question for the field profile of a conventional superconducting NMR magnet is about 35 cm,[41, 45] constant shuttling rates of ca. 0.25 T/ms were assumed.

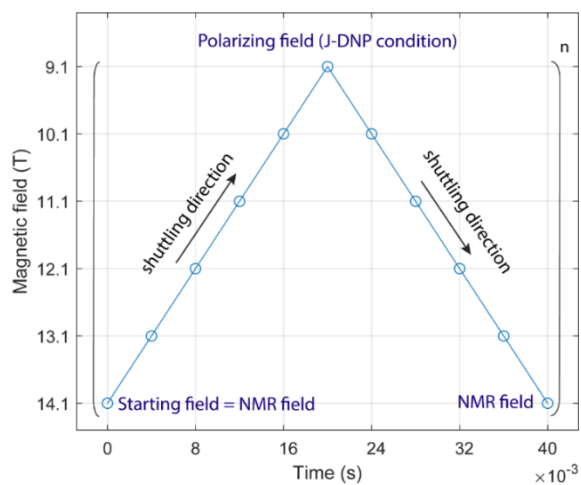

**FIG. 2.** Schematic description of the high→low→high B$_0$-cycling in MF-JDNP, in which the sample is shuttled n times at constant 0.25 T/ms rates from a starting magnetic field to a lower field corresponding to the JDNP condition, and then back to the NMR field in which the measurement is performed.

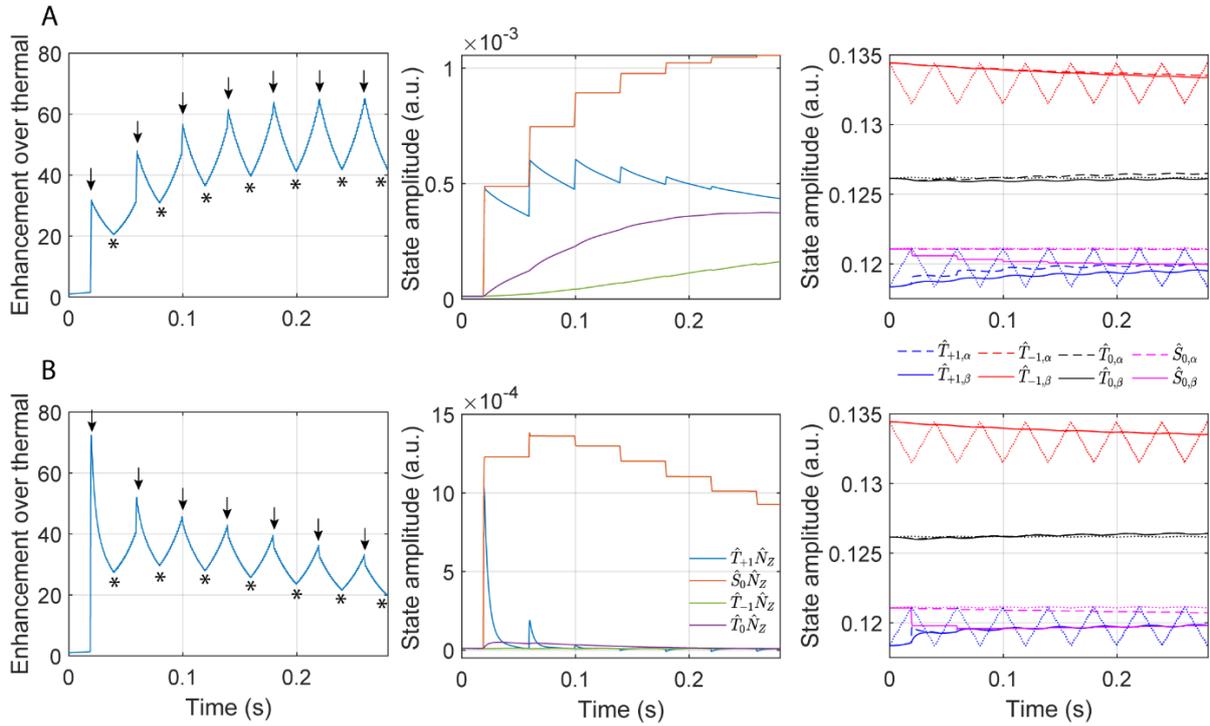

**FIG. 3:** MF-JDNP performed according to the scheme in Fig. 2, with n=7 loops between an NMR field of 14.1 T, and a 9.1 T polarizing field fulfilling the JDNP condition. (A) Simulations assuming a three-spin system where the solvent proton exchanges between configurations "A" and "B" (Fig. 1). (B) Simulations assuming that the solvent proton was fixed in configuration "A". In neither case were intra-radical protons positioned in "C" considered. *Left-hand column*: Time/magnetic field evolution of the nuclear enhancement, scaled over the thermal equilibrium value at each magnetic field. The black arrows indicate the JDNP condition; asterisks indicate NMR observation points. *Center column:* Time/magnetic field evolution of the $\hat{S}_0\hat{N}_z$, $\hat{T}_0\hat{N}_z$ and $\hat{T}_\pm\hat{N}_z$ states arising from the population imbalance between the α and β nuclear components of the singlet and triplet states; the sum of all these states corresponds to nuclear polarization scaled over the thermal equilibrium value at each magnetic field shown in the left-hand column. *Right-hand column:* Time/magnetic field evolution of the $\hat{S}_{0,\alpha/\beta}$, $\hat{T}_{0,\alpha/\beta}$ and $\hat{T}_{\pm,\alpha/\beta}$ states (straight/dashed lines, respectively) compared to their thermal equilibrium values (dotted lines, the α and β nuclear components are overlapped).

Figures 3A and 3B show the consequence of the ensuing shuttling on the nuclear polarization for a three-spin system with parameters as given in Table 1, with and without assuming exchanges between proton configurations "A" and "B", respectively. The left-hand column in Fig. 3 shows the enhancements over the thermal nuclear polarization that will be achieved in each case from such experiment; the center column clarifies this further, by showing the fate of the various spin states that add up to the total nuclear polarization $\hat{N}_z$ throughout the process. The right-hand column summarizes the physics of these events, by depicting the differential behavior of the various $\hat{T}_{\pm1,\beta/\alpha}$, $\hat{T}_{0,\beta/\alpha}$ and $\hat{S}_{0,\beta/\alpha}$ operators describing the triplet and singlet electronic coupled to the β/α nuclear spin states.

At the crux of the proposal lie shuttling speeds that, even if leading to magnetic field change rates that are still considerably slower than the Redfield relaxation rates of the electronic triplet states, are sufficiently fast for taking these states slightly out of the thermal equilibrium. These perturbations are illustrated in the right-hand column of Fig. 3, which compares the actual values of the above-mentioned states (with the straight and dashed lines representing the nuclear β/α states) vs their thermal equilibrium values (dotted lines). As these perturbed systems reach the JDNP condition, spectral densities reignite cross-relaxation processes between the $\hat{T}_{+1,\beta/\alpha}$ and $\hat{S}_{0,\beta/\alpha}$ states, which at other fields were too inefficient due to the large energy gap introduced by $J_{ex}$. The rates of these cross-relaxation processes depend on the $^1$H spin state, resulting in a temporary imbalance between the $\alpha$ and $\beta$ nuclear components associated to the triplet and –in particular– to the singlet states. The process that leads to this imbalance can be appreciated from Eqs. S1-S5 in the Supporting Information, which represent the relaxation rates of the of the $\alpha$ and $\beta$ nuclear components of the electronic singlet and triplet states. These equations predict that for the symmetric biradicals being here considered, singlet self-relaxation rates will be dominated by the $\left(\Delta^2_{\mathrm{AHF}}/30\right)J\left(J_{ex}+\omega_E+\omega_N\right)$ term for $\hat{S}_{0,\beta}$, and by $\left(\Delta^2_{\mathrm{AHF}}/30\right)J\left(J_{ex}-\omega_E-\omega_N\right)$ for $\hat{S}_{0,\alpha}$, where the $\Delta^2_{\mathrm{AHF}}$ is the second rank norm squared arising from anisotropies associated to the difference between the dipolar hyperfine coupling tensors between the protons and the two electrons, and $\omega_E$, $\omega_N$ are the electron and the nuclear Larmor frequencies. A negative $J_{ex} \approx \omega_E$ thus leads to $-R\left[\hat{S}_{0,\beta}\right] \gg -R\left[\hat{S}_{0,\alpha}\right]$, while a positive $J_{ex} \approx -\omega_E$ leads to $-R\left[\hat{S}_{0,\alpha}\right] \gg -R\left[\hat{S}_{0,\beta}\right]$. In either case a difference between the self-relaxation rates of $\hat{S}_{0,\alpha}$ and $\hat{S}_{0,\beta}$ leads to a population imbalance – and hence the creation of a transient, net nuclear magnetization enhancement. Such imbalance is reflected by the creation of $\hat{S}_0\hat{N}_Z = \hat{S}_{0,\alpha} - \hat{S}_{0,\beta}$, $\hat{T}_0\hat{N}_Z = \hat{T}_{0,\alpha} - \hat{T}_{0,\beta}$ and $\hat{T}_\pm\hat{N}_Z = \hat{T}_{\pm,\alpha} - \hat{T}_{\pm,\beta}$ states (Fig. 3, central column), and therefore by an overall nuclear magnetization enhancement given by $\hat{N}_Z = \left(\hat{S}_0\hat{N}_Z + \hat{T}_0\hat{N}_Z + \hat{T}_\pm\hat{N}_Z\right)/2$.

The enhancement predicted by this repeated shuttling is relatively isotropic [30]; it is maximized each time the JDNP condition is fulfilled, but begins to decay as the sample departs from this condition. This explains the oscillations displayed by $\hat{N}_Z$ with the shuttling; oscillations which are magnified further when considering the equilibrium nuclear polarization at each field (Fig. 3, left-hand column). Still, as the field where the JDNP condition is maximal will in general not correspond with a traditional NMR observation field, the MF-JDNP approach assumes an additional shuttling back to the homogeneous 14.1 T field region for a conventional NMR observation. While lowering the enhancement that could be achieved if remaining at the JDNP condition, a significant NMR enhancement is still predicted. It is also enlightening to compare Fig. 3A, which assumes that the nuclear spin can diffuse in-and-out of the polarization sphere, with Fig. 3B which assumes the spin spends all of its time in the polarizing "A" region. In the latter case, the shuttling leads to a clearly higher initial JDNP effect; however, the faster spin relaxation characterizing these electron-proximate nuclei, also leads to a rapid loss of this nuclear enhancement as the sample travels to the NMR-detection field. By contrast, the buildup in the former case is slower, but builds up to higher final values upon looping. It appears, therefore that diffusive processes end up having positive effects on the proposed scheme.

The aforementioned predictions assumed a three-spin system; Figure 4A shows the expectations arising from MF-JDNP if considering a four-spin system, which includes the presence of an intra-radical proton residing (without exchange) in region "C", that is dipole- and scalar-coupled to the electrons. The addition of this 4$^{th}$ spin will decrease the enhancement of the "solvent" $^1$H by ca. an order of magnitude, as most of the electron polarization imbalance created by the shuttling is now captured by the proton that's closer to the biradical. At the same time, this ca. 80-fold polarization enhancement of the intra-radical $^1$H will also be lost quickly, due to the high self-relaxation rates induced by the nearby electrons. On the other hand, replacing the intra-radical proton by deuterium leads to an increase of the radical's $T_1$ [57] and reinstates a sizable enhancement (Fig. 4B, blue line) – even if it is still ≈40% smaller than in the absence of any radical-based nucleus.

A final, important ingredient that may define the success of failure of the MF-JDNP strategy, concerns the presence of additional electronic spin-lattice relaxation processes; in particular, of relaxation mechanisms other than those treated by Redfield model. For example, both for the case of trityl- and of nitroxide-based monoradicals (and presumably for their biradicals as well), vibrational modes coupling the spins with orbital angular momenta fluctuations, are known to lead to substantial decreases in the electron spin relaxation times.[57] The magnitude of these relaxation rates can be high, reaching into ~$10^4$ and ~$10^6$ Hz for trityls and nitroxides, respectively.[58-60] Further, these effects are at the moment virtually impossible to calculate accurately from first principles –even if they can be inferred from vibrational measurements. These vibrational modulations may or may not interact with a biradical's singlet state, but will in all likelihood lead to significant changes in the $\hat{T}_{\pm,\alpha/\beta}$'s relaxation times, bringing them down to the ≈µs range (see Supporting Information for more details). This could profoundly affect the MF-JDNP experiment, as illustrated in Fig. 4B for the case of a four-spin system that is now affected by vibrations-driven relaxation modes whose magnitude were estimated based on reports for monotrityl radicals.[57, 59, 60] Not surprisingly, the addition of such strong competing relaxation mechanism will cancel out almost entirely the polarization enhancement effects in the solvent expected from the MF-JDNP methodology (Fig. 4B, violet trace).

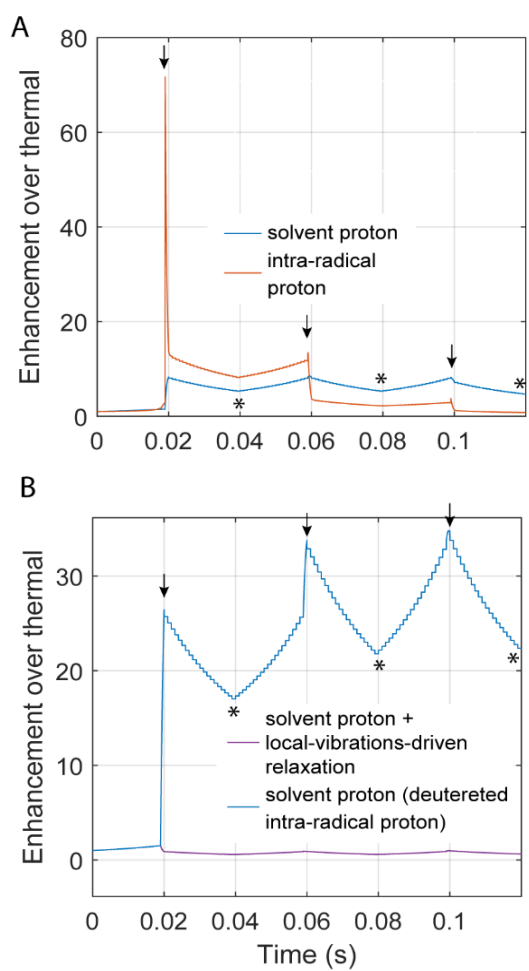

FIG. 4: Expectations of MF-JDNP experiments performed according to the scheme in Fig. 2, with three high-low-high $B_0$ shuttling repetitions. All plots show time/magnetic field evolution of the nuclear enhancement over the thermal equilibrium value. The black arrows indicate the JDNP condition, asterisks indicate potential NMR observation points. (A) Predictions for a four-spin system including a fixed intra-radical proton (the electron - proton scalar coupling was set to 1 MHz) and a diffusion "solvent". (B) Idem, but after replacing the intra-radical proton by a deuterium (blue line) but including an *ad hoc* term in the Redfield relaxation superoperator, applied only to the electron longitudinal states, representing a 6x$10^4$ Hz local-vibrations-driven contribution to the relaxation modes (violet line).

## IV. Discussion and Conclusions

This study explored the possibility of combining the JDNP effect that will spontaneously transfer electron polarization to nearby nuclei under $J_{ex}≈±ω_E$ conditions, with rapid sample field cycling. Exchange couplings in the order of several GHz have been reported for a number of biradicals using monophenyl, biphenyls and acetylene linkers.[39, 61-63] Pure hydrocarbon biradicals created using these linkers are expected to be conformationally rigid,[63] leading to a $J_{ex}$ value that will remain constant during the JDNP nuclear polarization build-up (while modulation of $J_{ex}$ due to putative conformational dynamics will not lead to shortening in the electrons state relaxation rates), thereby enabling the JDNP experiment.[64] The question now is how to use JDNP's potential, for enhancing NMR sensitivity at high magnetic fields without microwave irradiation. MF-JDNP could achieve this, by exploiting shuttling rates of ≈T/ms in order to create a sufficient disturbance in the electron polarization; only under such conditions, will an out-of-equilibrium situation be created in the absence of microwaves. Then, as the sample is cycled through fields that include fulfillment of the JDNP condition, nuclear polarization is spontaneously created. The present study assumed a lower-field JDNP condition and shuttling back to higher field for NMR measurements; alternatives providing comparable nuclear enhancements while shuttling from lower NMR to higher JDNP-fulfilling fields, can also be devised. All the scenarios that were here analyzed involved radicals whose electron relaxation times were dominated by dipolar hyperfine relaxation, and singlet and triplet's $T_1$s comparable to the shuttling times. Notice that, as further discussed in the Supporting Information, these triplet and singlet relaxation rates can be orders-of-magnitude smaller than longitudinal $T_1$ electron relaxation rates –which reach in excess of ≈$10^6$ Hz in biradicals at any magnetic field.[59] The presence of intra-radical protons can affect these rates and decrease the nuclear hyperpolarization of the solvent; however, this can be largely restored if the former are substituted by deuterons. Eventually, however, the presence of very strong competing relaxation mechanism like those stemming from local vibrational modes – in the case of trityls, arising from the stretching of the C-S bond in trityl radicals – might shorten further the lifetimes of the above-mentioned states, eliminating the MF-JDNP effect altogether. These effects will arise from the mixing between spin and orbital angular momenta, as driven by spin-orbit coupling (SOC). Although detrimental for MF-JDNP, these SOCs can be suppressed by eliminating the heavy atoms (heavier than $^{19}$F) from the biradical structure, thereby restoring sufficiently slow electron relaxation rates to support MF-JDNP. [55, 65] Alternatives to bypass such electron relaxation mechanisms competing with Redfield relaxation are also important in spintronics, and therefore actively being sought.[55] Additional electron relaxation mechanisms might arise due to vibrations and collisions with surrounding diamagnetic molecules;[66] however, these processes are not expected in biradicals that make weak intermolecular interactions with the solvent. The MF-JDNP experiment might thus be realized using deuterated carbon-centered hydrocarbon radical centers free from heteroatoms, linked by mono-phenyl or acetylene units.[63] From an instrumentation standpoint current shuttling technologies could enable field disturbances on the order of ~0.25 T/ms, [41, 45] sufficient to enable the MF-JDNP effect. Tests based on these chemical and technological systems are currently in progress.

## V. Acknowledgments

This project was funded by the Israel Science Foundation (ISF 1874/22), the Minerva Foundation (Germany), and the US National Science Foundation (grants number CHE-2203405). MGC acknowledges Weizmann's Faculty of Chemistry for a Dean Fellowship. LF holds the Bertha and Isadore Gudelsky Professorial Chair and Heads the Clore Institute for High-Field Magnetic Resonance Imaging and Spectroscopy whose support is acknowledged, as is the generosity of the Perlman Family Foundation. The authors acknowledge Profs. Ilya Kuprov, Olav Schiemann and Mr Kevin Kopp for discussions.

**Supplementary material for**

**Microwave-free J-driven DNP (MF-JDNP): A proposal for enhancing the sensitivity of solution-state NMR**

Maria Grazia Concilio and Lucio Frydman

*Department of Chemical and Biological Physics, Weizmann Institute of Science, Rehovot, Israel*

**Contents**

## A. Spin relaxation and proton random walk codes used in the present analysis

The following codes were used to carry out the computations described in the main text of this study:

1) Code used to perform simulations shown in Figure 3 of the main text, to be run on Spinach version 2.7.6049 or later.

```matlab
% MFJDNP with fixed proton position.
%
% maria-grazia.concilio@weizmann.ac.il

function MFJDNP_nodynamics()

%Spin system
sys.isotopes={'1H','E','E'};

% Carbon coordinates
ta_sum=[0,0,-7.20]';
tb_sum=[0,0,7.20]';

% g-tensor tensor
g_tensor=[2.0032,0,0;0,2.0032,0;0,0,2.0026];
g_tensor1=[2.0032,0,0;0,2.0032,0;0,0,2.0026];

% Nuclear Zeeman interaction (for DFT for proton)
inter.zeeman.matrix{1}=[5,0,0;0,5,0;0,0,5];
inter.zeeman.matrix{2}=axis_tsymm(g_tensor,ta_sum);
inter.zeeman.matrix{3}=axis_tsymm(g_tensor1,tb_sum);

% Exchange
inter.coupling.scalar=cell(3,3);
measurement_field=9.1;
inter.coupling.scalar{2,3}=((measurement_field*spin('E')/(2*pi))+ ...
                            (measurement_field*spin('1H')/(2*pi)));

% Basis set
bas.formalism='sphten-liouv';
bas.approximation='none';

% Relaxation theory
inter.relaxation={'redfield','SRFK'}; % Scalar
relaxationinter.equilibrium='dibari';
inter.rlx_keep='labframe';
inter.temperature=298;
inter.tau_c={2200e-12};
sys.tols.rlx_integration=1e-12; % Needs to be this tight
inter.srfk_assume='labframe';
inter.srfk_tau_c={[1.0 1e-12]};
inter.srfk_mdepth=cell(3);
inter.srfk_mdepth{2,3}=1e9;
sys.output='hush';

% Get the initial state at the thermal equilibrium
sys.magnet=14.1;

% parameters
starting_field=sys.magnet;
final_field=9.1;
field_gap=0.25;

% parameters
field_ramp=final_field:field_gap:starting_field;
field_grid_up=flip(field_ramp);
field_grid_down=flip(field_grid_up);
field_grid_down=field_grid_down(2:end-1);
field_grid=[field_grid_up(2:end) field_grid_down starting_field ...
            field_grid_up(2:end) field_grid_down starting_field ...
            field_grid_up(2:end) field_grid_down starting_field];

% Get thermal equilibrium at starting field
spin_system=create(sys,inter);
spin_system=basis(spin_system,bas);
spin_system=assume(spin_system,'labframe');
H0=hamiltonian(spin_system,'left');
rho_eq=equilibrium(spin_system,H0);

% Set loops
dt=1e-5;

%Time axis generation
field_steps=400;

%Time axis generation
time_axis=linspace(0,((field_steps/4)*dt)*numel(field_grid), ...
                   ((field_steps/4)*numel(field_grid)));

% Preallocate the trajectory
traj=[]; sing_trip_eqa=[];

for n=1:numel(field_grid)

    % Magnet
    sys.magnet=field_grid(n);
    disp(['n = ' num2str(n) ...
          ', field = ' num2str(sys.magnet)]);

    % Set current coordinate
    inter.coordinates={[1.266,1.606,2.255];
                       [0,0,-7.20];
                       [0,0,+7.20]};
    spin_system=create(sys,inter);
    spin_system=basis(spin_system,bas);
    R3=relaxation(spin_system);
    H3=hamiltonian(assume(spin_system,'labframe'));
    traj3=evolution(spin_system,H3+1i*R3,[],rho_eq,...
                    dt,99,'trajectory');
    traj=[traj traj3]; %#ok<AGROW>
    rho_eq=traj3(:,end);

    % Get thermal equilibrium
    rhoeq2=equilibrium(basis(create(sys,inter),bas),hamiltonian(assume(basis(create(sys,inter),bas),'labframe'),'left'));
    drho2=rhoeq2.*ones(size(traj3));
    sing_trip_eqa=[sing_trip_eqa drho2]; %#ok<AGROW>
end

% Build the component operators
unit=state(spin_system,{'E','E','E'},{1,2,3},'exact');
Mz=state(spin_system,{'Lz','E','E'},{1,2,3},'exact');
Lz=state(spin_system,{'E','Lz','E'},{1,2,3},'exact');
Sz=state(spin_system,{'E','E','Lz'},{1,2,3},'exact');
MzLz=state(spin_system,{'Lz','Lz'},{1,2},'exact');
MzSz=state(spin_system,{'Lz','Lz'},{1,3},'exact');
LzSz=state(spin_system,{'Lz','Lz'},{2,3},'exact');
LmSp=state(spin_system,{'L-','L+'},{2,3},'exact');
LpSm=state(spin_system,{'L+','L-'},{2,3},'exact');
MzLzSz=state(spin_system,{'Lz','Lz','Lz'},{1,2,3},'exact');
MzLmSp=state(spin_system,{'Lz','L-','L+'},{1,2,3},'exact');
MzLpSm=state(spin_system,{'Lz','L+','L-'},{1,2,3},'exact');

% Build the alpha singlet and triplet states
S0a=(unit/8)+(Mz/4)-(LmSp/4)-(LpSm/4)-(LzSz/2)-(MzLmSp/2)-(MzLpSm/2)-MzLzSz;
S0b=(unit/8)-(Mz/4)-(LmSp/4)-(LpSm/4)-
(LzSz/2)+(MzLmSp/2)+(MzLpSm/2)+MzLzSz;
T0a=(unit/8)+(Mz/4)+(LmSp/4)+(LpSm/4)-(LzSz/2)+(MzLmSp/2)+(MzLpSm/2)-MzLzSz;
T0b=(unit/8)+(Mz/4)+(LmSp/4)+(LpSm/4)-(LzSz/2)-(MzLmSp/2)-
(MzLpSm/2)+MzLzSz;
Tpa=(unit/8)+(Mz/4)+(Lz/4)+(Sz/4)+(LzSz/2)+(MzSz/2)+(MzLz/2)+MzLzSz;
Tpb=(unit/8)-(Mz/4)+(Lz/4)+(Sz/4)+(LzSz/2)-(MzSz/2)-(MzLz/2)-MzLzSz;
Tma=(unit/8)+(Mz/4)-(Lz/4)-(Sz/4)+(LzSz/2)-(MzSz/2)-(MzLz/2)+MzLzSz;
Tmb=(unit/8)-(Mz/4)-(Lz/4)-(Sz/4)+(LzSz/2)+(MzSz/2)+(MzLz/2)-MzLzSz;

% Build the Nz-singlet and Nz-triplet states
SNz=(Mz/4)-(MzLmSp/2)-(MzLpSm/2)-MzLzSz; SNz=SNz/norm(SNz,2);
TpNz=(Mz/4)+(MzLz/2)+(MzSz/2)+MzLzSz; TpNz=TpNz/norm(TpNz,2);
TmNz=(Mz/4)-(MzLz/2)-(MzSz/2)+MzLzSz; TmNz=TmNz/norm(TmNz,2);
T0Nz=(Mz/4)+(MzLmSp/2)+(MzLpSm/2)-MzLzSz; T0Nz=T0Nz/norm(T0Nz,2);

% Project out the observables
S0a1=real(S0a'*traj); S0b1=real(S0b'*traj);
T0a1=real(T0a'*traj); T0b1=real(T0b'*traj);
Tpa1=real(Tpa'*traj); Tpb1=real(Tpb'*traj);
Tma1=real(Tma'*traj); Tmb1=real(Tmb'*traj);
TpNz1=real(TpNz'*traj); SNz1=real(SNz'*traj);
TmNz1=real(TmNz'*traj); T0Nz1=real(T0Nz'*traj);
S0a_eq=real(S0a'*sing_trip_eqa); S0b_eq=real(S0b'*sing_trip_eqa);
T0a_eq=real(T0a'*sing_trip_eqa); T0b_eq=real(T0b'*sing_trip_eqa);
Tpa_eq=real(Tpa'*sing_trip_eqa); Tpb_eq=real(Tpb'*sing_trip_eqa);
Tma_eq=real(Tma'*sing_trip_eqa); Tmb_eq=real(Tmb'*sing_trip_eqa);

% Get Nz
Nztot=state(spin_system,'Lz','1H'); Nztot=Nztot/norm(Nztot,2);
Nztot_1=real(Nztot'*traj);
Nztot_eq=real(Nztot'*sing_trip_eqa);
Enhanc_nztot=Nztot_1./Nztot_eq;

% Do plotting
figure();
subplot(1,3,1);
plot(time_axis,Enhanc_nztot); hold on; grid on;
xlabel('Time (s)'); ylabel('Enhancement over thermal');
subplot(1,3,2);
plot(time_axis,Tpa1,'b--',time_axis,Tpb1,'b-',time_axis,Tma1,'r--',...
     time_axis,Tmb1,'r-', ...
     time_axis,T0a1,'k--',time_axis,T0b1,'k-',time_axis,S0a1,'m--',...
     time_axis,S0b1,'m-'); grid on; hold on;
xlabel('Time (s)'); ylabel('State amplitude (a.u.)');
legend({'${\hat T _ {+1,\alpha}}$','${\hat T _ {+1,\beta}}$', ...
        '${\hat T _ {-1,\alpha}}$','${\hat T _ {-1,\beta}}$' ...
        '${\hat T _ {0,\alpha}}$','${\hat T _ {0,\beta}}$' ...
        '${\hat S _ {0,\alpha}}$','${\hat S _ {0,\beta}}$'},...
        'Interpreter','latex','Location','northeast'); legend boxoff
ylabel('State amplitude (a.u.)');
plot(time_axis,Tpa_eq,'b:',time_axis,Tpb_eq,'b:',time_axis,Tma_eq,'r:',time_axis,Tmb_eq,'r:', ...
     time_axis,T0a_eq,'k:',time_axis,T0b_eq,'k:',time_axis,S0a_eq,'m:',time_axis,S0b_eq,'m:'); hold on;
subplot(1,3,3);
plot(time_axis,TpNz1/2,time_axis,SNz1/2,time_axis,TmNz1/2,time_axis,T0Nz1/2); hold on;
grid on;
xlabel('Time (s)');
legend({'${\hat T _ {+1}\hat N _ {Z}}$', ...
        '${\hat S _ {0}\hat N _ {Z}}$', ...
        '${\hat T _ {-1}\hat N _ {Z}}$', ...
        '${\hat T _ {0}\hat N _ {Z}}$'},...
        'Interpreter','latex','Location','northeast'); legend boxoff

end
```

## 2) Code used to perform simulations shown in Figure 4 of the main text, to be run on Spinach version 2.7.6049 or later.

```matlab
% MFJDNP including exchange between the site A and the site B.
%
% maria-grazia.concilio@weizmann.ac.il

function MFJDNP_with_dynamics()

%Spin system
sys.isotopes={'1H','E','E','1H'};
% Note 1: replace the 4th spin 1H with 2H to consider deuterium
% Note 2: remove the 4th spin to obtain the result shown in Figure 3

% Set spin coordinates, the proton in the solvent was set in the site B
inter.coordinates={[15.457,10.052,-8.0];
                   [0,0,-7.20];
                   [0,0,+7.20];
                   [0,0,10.4]};

% Set g-tensors
Rot1=euler2dcm(0,0,0); Rot2=euler2dcm(0,0,0);
g_tensor1=Rot1*[2.0032,0,0;0,2.0032,0;0,0,2.0026]*Rot1';
g_tensor2=Rot2*[2.0032,0,0;0,2.0032,0;0,0,2.0026]*Rot2';

% Nuclear and electron Zeeman interactions
inter.zeeman.matrix{1}=[5,0,0;0,5,0;0,0,5];
inter.zeeman.matrix{2}=axis_tsymm(g_tensor1,[0,0,-7.20]');    % rotate tensor 1 to satisfy axial symmetry
inter.zeeman.matrix{3}=axis_tsymm(g_tensor2,[0,0,+7.20]');    % rotate tensor 2 to satisfy axial symmetry
inter.zeeman.matrix{4}=[5,0,0;0,5,0;0,0,5];

% Set exchange couling
inter.coupling.scalar=cell(4,4);
polarization_field=9.1;                % set polarization field
inter.coupling.scalar{2,3}=((polarization_field*spin('E')/(2*pi))+ ...
                           (polarization_field*spin('1H')/(2*pi)));
% % Set proton-electron scalar coupling for the intra-radical proton
% inter.coupling.scalar{3,4}=1e6;
%
% % Set quadrupoplar coupling when deuterium is present
% inter.coupling.matrix{4,4}=eeqq2nqi(1.18e6,0.53,1,[0 0 0]);

% Basis set
bas.formalism='sphten-liouv';
bas.approximation='none';

% Relaxation theory
inter.relaxation={'redfield','SRFK'}; % Scalar relaxation
inter.equilibrium='dibari';
inter.rlx_keep='labframe';
inter.temperature=298;
inter.tau_c={2200e-12};
sys.tols.rlx_integration=1e-12; % Needs to be this tight
inter.srfk_assume='labframe';
inter.srfk_tau_c={[1.0 1e-12]};
inter.srfk_mdepth=cell(4);
inter.srfk_mdepth{2,3}=1e9;
sys.output='hush';

% Set the initial magnetic field
sys.magnet=14.1;
starting_field=sys.magnet;

% Shuttlign rate
shuttling_rate=0.25; % T/ms

% Set magnetic field grid
field_ramp=polarization_field:shuttling_rate:starting_field;
field_grid_up=flip(field_ramp);
field_grid_down=flip(field_grid_up);
field_grid_down=field_grid_down(2:end-1);
field_grid=[field_grid_up(2:end) field_grid_down starting_field ...
            field_grid_up(2:end) field_grid_down starting_field ...
            field_grid_up(2:end) field_grid_down starting_field];

% Spinach housekeeping
spin_system=create(sys,inter);
spin_system=basis(spin_system,bas);

% Get thermal equilibrium at starting field
rho_eq=equilibrium(spin_system, ...
             hamiltonian(assume(spin_system,'labframe'),'left'));

% Set timings
st_conf_a=1e-6;   % time spent in the site A
st_conf_b=9e-6;   % time spent in the site B
field_steps=100;

%Time axis
time_axis=linspace(0,((field_steps)*(st_conf_a+st_conf_b))*numel(field_grid), ...
                     ((field_steps)*numel(field_grid)));

% Preallocate the trajectory
traj=[]; sing_trip_eqa=[];

% Loop over field grid
for n=1:numel(field_grid)

    % Set magnetic field
    sys.magnet=field_grid(n);
    disp(['n = ' num2str(n) ...
          ', field = ' num2str(sys.magnet)]);

    % Run with A <->B exchange at the matching conditon
    if sys.magnet == polarization_field

        % Let exchange occur for 1 ms
        for k=1:(field_steps)

            disp(['k = ' num2str(k)]);

            % Set spin coordinates, the proton in the solvent was set in the site A
            inter.coordinates={[1.266,1.606,2.255];
                               [0,0,-7.20];
                               [0,0,+7.20];
                               [0,0,10.4]};

            % Set Hamiltonian and relaxation superoperator
            spin_system=create(sys,inter);
            spin_system=basis(spin_system,bas);
            R=relaxation(spin_system);
            H=hamiltonian(assume(spin_system,'labframe'));

            % Spend 1e-6 s in the site A
            traj1=evolution(spin_system,H+1i*R,[],rho_eq,st_conf_a,1,'final');

            % Get thermal equilibrium (for normalization purposes)
            rhoeq1=equilibrium(basis(create(sys,inter),bas),hamiltonian(assume(basis(create(sys,inter),bas),'labframe'),'left'));
            drho1=rhoeq1.*ones(size(traj1));

            % Set spin coordinates, the proton in the solvent was set in the
            % site B
            inter.coordinates={[15.457,10.052,-8.0];
                               [0,0,-7.20];
                               [0,0,+7.20];
                               [0,0,10.4]};

            % Set Hamiltonian and relaxation superoperator
            spin_system=create(sys,inter);
            spin_system=basis(spin_system,bas);
            R1=relaxation(spin_system);
            H1=hamiltonian(assume(spin_system,'labframe'));

            % Spend 9e-6 s in the conformation B
            traj2=evolution(spin_system,H1+1i*R1,[],traj1(:,end),st_conf_b,1,'final');

            % Collect trajectory
            traj=[traj traj2]; %#ok<AGROW>
            rho_eq=traj2(:,end);

            % Collect thermal equilibrium trajectory
            sing_trip_eqa=[sing_trip_eqa drho1]; %#ok<AGROW>

        end

        % Run simulation in the configuration B till the matching condition is
        % satisfied
    else

        % Set spin coordinates, the proton in the solvent was set in the
        % site B
        inter.coordinates={[15.457,10.052,-8.0];
                           [0,0,-7.20];
                           [0,0,+7.20];
                           [0,0,10.4]};

        % Set Hamiltonian and relaxation superoperator
        spin_system=create(sys,inter);
        spin_system=basis(spin_system,bas);
        R3=relaxation(spin_system);
        H3=hamiltonian(assume(spin_system,'labframe'));

        % Let evolve for 1 ms
        traj3=evolution(spin_system,H3+1i*R3,[],rho_eq,...
                        1e-5,99,'trajectory');

        % Collect trajectory
        traj=[traj traj3]; %#ok<AGROW>
        rho_eq=traj3(:,end);

        % Get thermal equilibrium (for normalization purposes)
        rhoeq2=equilibrium(basis(create(sys,inter),bas),hamiltonian(assume(basis(create(sys,inter),bas),'labframe'),'left'));
        drho2=rhoeq2.*ones(size(traj3));

        % Collect thermal equilibrium trajectory
        sing_trip_eqa=[sing_trip_eqa drho2]; %#ok<AGROW>

    end
end

% Build the component operators
unit=state(spin_system,{'E','E','E'},{1,2,3},'exact');
Mz=state(spin_system,{'Lz','E','E'},{1,2,3},'exact');
Lz=state(spin_system,{'E','Lz','E'},{1,2,3},'exact');
Sz=state(spin_system,{'E','E','Lz'},{1,2,3},'exact');
MzLz=state(spin_system,{'Lz','Lz'},{1,2},'exact');
MzSz=state(spin_system,{'Lz','Lz'},{1,3},'exact');
LzSz=state(spin_system,{'Lz','Lz'},{2,3},'exact');
LmSp=state(spin_system,{'L-','L+'},{2,3},'exact');
LpSm=state(spin_system,{'L+','L-'},{2,3},'exact');
MzLzSz=state(spin_system,{'Lz','Lz','Lz'},{1,2,3},'exact');
MzLmSp=state(spin_system,{'Lz','L-','L+'},{1,2,3},'exact');
MzLpSm=state(spin_system,{'Lz','L+','L-'},{1,2,3},'exact');

% Build the alpha and beta singlet and triplet states
S0a=(unit/8)+(Mz/4)-(LmSp/4)-(LpSm/4)-(LzSz/2)-(MzLmSp/2)-(MzLpSm/2)-MzLzSz;
S0b=(unit/8)-(Mz/4)-(LmSp/4)-(LpSm/4)-(LzSz/2)+(MzLmSp/2)+(MzLpSm/2)+MzLzSz;
T0a=(unit/8)+(Mz/4)+(LmSp/4)+(LpSm/4)-(LzSz/2)-(MzLmSp/2)-(MzLpSm/2)-MzLzSz;
T0b=(unit/8)-(Mz/4)+(LmSp/4)+(LpSm/4)-(LzSz/2)-(MzLmSp/2)-(MzLpSm/2)+MzLzSz;
Tpa=(unit/8)+(Mz/4)+(Lz/4)+(Sz/4)+(LzSz/2)+(MzSz/2)+(MzLz/2)+MzLzSz;
Tpb=(unit/8)-(Mz/4)+(Lz/4)+(Sz/4)+(LzSz/2)-(MzSz/2)-(MzLz/2)-MzLzSz;
```

```matlab
Tma=(unit/8)+(Mz/4)-(Lz/4)-(Sz/4)+(LzSz/2)-(MzSz/2)-(MzLz/2)+MzLzSz;
Tmb=(unit/8)-(Mz/4)-(Lz/4)-(Sz/4)+(LzSz/2)+(MzSz/2)+(MzLz/2)-MzLzSz;

% Build the Nz-singlet and Nz-triplet states
SNz=(Mz/4)-(MzLmSp/2)-(MzLpSm/2)-MzLzSz; SNz=SNz/norm(SNz,2);
TpNz=(Mz/4)+(MzLz/2)+(MzSz/2)+MzLzSz; TpNz=TpNz/norm(TpNz,2);
TmNz=(Mz/4)-(MzLz/2)-(MzSz/2)+MzLzSz; TmNz=TmNz/norm(TmNz,2);
T0Nz=(Mz/4)+(MzLmSp/2)+(MzLpSm/2)-MzLzSz; T0Nz=T0Nz/norm(T0Nz,2);

% Project out the observables
S0a1=real(S0a'*traj); S0b1=real(S0b'*traj);
T0a1=real(T0a'*traj); T0b1=real(T0b'*traj);
Tpa1=real(Tpa'*traj); Tpb1=real(Tpb'*traj);
Tma1=real(Tma'*traj); Tmb1=real(Tmb'*traj);
TpNz1=real(TpNz'*traj); SNz1=real(SNz'*traj);
TmNz1=real(TmNz'*traj); T0Nz1=real(T0Nz'*traj);

% Project out thermal equilibrium values
S0a_eq=real(S0a'*sing_trip_eqa); S0b_eq=real(S0b'*sing_trip_eqa);
T0a_eq=real(T0a'*sing_trip_eqa); T0b_eq=real(T0b'*sing_trip_eqa);
Tpa_eq=real(Tpa'*sing_trip_eqa); Tpb_eq=real(Tpb'*sing_trip_eqa);
Tma_eq=real(Tma'*sing_trip_eqa); Tmb_eq=real(Tmb'*sing_trip_eqa);

% Get Nz trajectories
Nz1=state(spin_system,{'Lz'},{1}); Nz1=Nz1/norm(Nz1,2);
Nz1_1=real(Nz1'*traj);
Nz1_eq=real(Nz1'*sing_trip_eqa);
Nz4=state(spin_system,{'Lz'},{4}); Nz4=Nz4/norm(Nz4,2);
Nz4_1=real(Nz4'*traj);
Nz4_eq=real(Nz4'*sing_trip_eqa);

% Get enhancement
Enhanc1=Nz1_1./Nz1_eq;
Enhanc4=Nz4_1./Nz4_eq;

figure(2);
subplot(1,3,1);
plot(time_axis,Enhanc1,time_axis,Enhanc4); hold on; grid on;
xlabel('Time (s)'); ylabel('State amplitude (a.u.)');
subplot(1,3,2);
plot(time_axis,Tpa1,'b--',time_axis,Tpb1,'b-',time_axis,Tma1,'r--',time_axis,Tmb1,'r-', ...
    time_axis,T0a1,'k--',time_axis,T0b1,'k-',time_axis,S0a1,'m--',time_axis,S0b1,'m-'); grid on; hold on;
xlabel('Time (s)'); ylabel('State amplitude (a.u.)');
plot(time_axis,Tpa_eq,'b:',time_axis,Tpb_eq,'b:',time_axis,Tma_eq,'r:',time_axis,Tmb_eq,'r:', ...
    time_axis,T0a_eq,'k:',time_axis,T0b_eq,'k:',time_axis,S0a_eq,'m:',time_axis,S0b_eq,'m:'); hold on;
subplot(1,3,3);
plot(time_axis,TpNz1/2,time_axis,SNz1/2,time_axis,TmNz1/2,time_axis,T0Nz1/2); hold on; grid on;
xlabel('Time (s)'); ylabel('State amplitude (a.u.)');
legend({'${\hat T _ {+1}\hat N _ {Z}}$', ...
        '${\hat S _ {0}\hat N _ {Z}}$', ...
        '${\hat T _ {-1}\hat N _ {Z}}$', ...
        '${\hat T _ {0}\hat N _ {Z}}$'}, ...
        'Interpreter','latex','Location','northeast'); legend boxoff

end
```

## 3) Code written to the simulate a polarized proton random walk

```matlab
% maria-grazia.concilio@weizmann.ac.il

function proton_random_walk()

clear; close all;

% Set number of runs
run_number=10;

% Number of steps
nsteps=10000;

% Set inter-electron distance
distance=30; % Angstroms

for n=1:run_number

% Make a sphere
[x0,y0,z0]=sphere;

% Scale to desired radius
radius=4.5;
x1=x0*radius;
y1=y0*radius;
z1=z0*radius;

% Translate sphere to a location
offsetz1=-7.20; offsetz2=7.20;
offsety1=0; offsety2=0;
offsetx1=0; offsetx2=0;
zb=z1+offsetz1; za=z1+offsetz2;
yb=y1+offsety1; ya=y1+offsety2;
xb=x1+offsetx1; xa=x1+offsetx2;

% Set axes
xmin=min(min(xb))-(distance/2); xmax=max(max(xa))+(distance/2);
ymin=min(min(yb))-(distance/2); ymax=max(max(ya))+(distance/2);
zmin=min(min(zb))-(distance/2); zmax=max(max(za))+(distance/2);

% Set starting coordinates
x=xmin+(xmax-xmin)*rand(1,1);
y=ymin+(ymax-ymin)*rand(1,1);
z=zmin+(zmax-zmin)*rand(1,1);
if   ((x-offsetx1)^2+(y-offsety1)^2+(z-offsetz1)^2<=radius^2)||...
     ((x-offsetx2)^2+(y-offsety2)^2+(z-offsetz2)^2<=radius^2)
    error('Too close to the electron, run again!');
end

% Draw polarization sphere
figure(1);
h1=surf(x1,y1,za); hold on;
set(h1,'FaceColor',[0.5 0.5 0.5],'facealpha',0.5,'edgecolor','none');
line([0 0],[0 0],[-2.7 2.7],'color',[0.5 0.5 0.5]);
h2=surf(x1,y1,zb); hold on;
set(h2,'FaceColor',[0.5 0.5 0.5],'facealpha',0.5,'edgecolor','none');
plot3(x,y,z,'r*');
xlabel('X (Angstroms)'); ylabel('Y (Angstroms)'); zlabel('Z (Angstroms)');
axis equal;
axis([xmin xmax ymin ymax zmin zmax]);
radius1=10;
x2=x0*radius1+offsetx1; x3=x0*radius1+offsetx2;
y2=y0*radius1+offsetx1; y3=y0*radius1+offsety2;
z2=z0*radius1+offsetz1; z3=z0*radius1+offsetz2;
h3=surf(x2,y2,z2); hold on;
set(h3,'FaceColor','b','facealpha',0.05,'edgecolor','none');
h4=surf(x3,y3,z3); hold on;
set(h4,'FaceColor','b','facealpha',0.05,'edgecolor','none');

% Preallocate trajectory
trajectory=[]; i=0;

% Go with loop
while i<nsteps

    % Get a random direction
    direction=randi(6);

    dx=0;
    dy=0;
    dz=0;

    % Move of 1 Ang in any direction
    switch direction
        case 1
            dy=1;
        case 2
            dy=-1;
        case 3
            dx=1;
        case 4
            dx=-1;
        case 5
            dz=1;
        case 6
            dz=-1;
    end

    % Move forward
    nx=x+dx;
    ny=y+dy;
    nz=z+dz;

    % Avoid crossign spheres
    if   ((nx-offsetx1)^2+(ny-offsety1)^2+(nz-offsetz1)^2<=radius^2)||...
         ((nx-offsetx2)^2+(ny-offsety2)^2+(nz-offsetz2)^2<=radius^2)

        switch direction
            case 1
                ny=-1+(-1+ny);
            case 2
                ny=+1+(+1+ny);
            case 3
                nx=-1+(-1+nx);
            case 4
                nx=+1+(+1+nx);
            case 5
                nz=-1+(-1+nz);
            case 6
                nz=+1+(+1+nz);
        end
    end

    % Set periodic boundary conditions
    if nx<xmin
        x=xmax-(x+xmax);
        nx=xmax-(nx+xmax);
```

```matlab
        elseif nx>xmax
            x=xmax-(x-xmax);
            nx=xmax-(nx-xmax);
        end
        if ny<ymin
            y=ymax-(y+ymax);
            ny=ymax-(ny+ymax);
        elseif ny>ymax
            y=ymax-(y-ymax);
            ny=ymax-(ny-ymax);
        end
        if nz<zmin
            z=zmax-(z+zmax);
            nz=zmax-(nz+zmax);
        elseif nz>zmax
            z=zmax-(z-zmax);
            nz=zmax-(nz-zmax);
        end
        
        % Make trajectory
        trajectory=[trajectory;nx ny nz;
                    offsetx1 offsety1 offsetz1;...
                    offsetx2 offsety2 offsetz2]; %#ok<AGROW>
        
        % Move forward
        i=i+1;
        
        % Display step
        disp(['step number = ' num2str(i)]);
        
        % Do the drawing
%         line([x nx],[y ny],[z nz],'color','r');
%         plot3(nx,ny,nz,'r*','MarkerSize',2); hold on;
        x=nx; y=ny; z=nz;
        
    end
    
    traj_slices=permute(reshape(trajectory',[3,3,size(trajectory,1)/3]),[2,1,3]);
    pol_sphere=[]; out_pol_sphere=[];
    
    for k=1:(size(trajectory,1)/3)
        traj_slice=traj_slices(:,:,k);
        
        %current coordinate
        coordinates={(traj_slice(1,:))};
        xp=coordinates{1,1}(1); yp=coordinates{1,1}(2); zp=coordinates{1,1}(3);
        
        if  ((xp-offsetx1)^2+(yp-offsety1)^2+(zp-offsetz1)^2<=radius1^2)||...
                ((xp-offsetx2)^2+(yp-offsety2)^2+(zp-offsetz2)^2<=radius1^2)
            
            pol_sphere=[pol_sphere coordinates]; %#ok<AGROW>
        else
            out_pol_sphere=[out_pol_sphere coordinates]; %#ok<AGROW>
        end
    end
    
    % Display result
    disp({'Run = ' numel(n); 'In = ' numel(pol_sphere); ' Out =' numel(out_pol_sphere)});
    
    if ~isempty(pol_sphere)
        
    % Get new trajectory
    pol_sphere=pol_sphere';
    pol_sphere_mean=mean(cell2mat(pol_sphere));
    
    traj{1,1}=trajectory;
    run.trajecory(n)=traj;
    
    pol{1,1}=pol_sphere;
    run.pol_sphere(n)=pol;
    run.n_pol_sphere(n)=numel(pol_sphere);
    
    out_pol{1,1}=out_pol_sphere;
    run.out_pol_sphere(n)=out_pol;
    run.n_out_pol_sphere(n)=numel(out_pol_sphere);
    
    pol_mean{1,1}=pol_sphere_mean;
    run.mean(n)=pol_mean;
    
    end
end

% Save result
run.mean_pol_sphere=mean(run.n_pol_sphere);
run.mean_out_pol_sphere=mean(run.n_out_pol_sphere);
run.mean_coordinates=mean(cell2mat(run.mean'));
save('test.mat','run');

end
```

## B. Singlet and triplet relaxation rates for the intra-radical protons

We considered it useful to have an estimation of the relaxation rates of the triplet and singlet states, as a number of both the "radical" protons, and in the presence/absence of a local relaxation mechanism like that produced by molecular vibrations. To this end we start from the rates predicted by the Redfield model [30] as a function of the magnetic field; approximate rates for the electronic spin states can then be obtained by multiplying the hyperfine component of the Redfield relaxation rates by as many protons as are present in the radical, plus the addition of the local relaxation term. In this approximation, one also assumes negligible scalar hyperfine interactions, comparable hyperfine couplings between the protons within a single radical unit and their closest electron, and negligible hyperfine couplings between these protons and their more distant electron. The ensuing analytical expressions for the relaxation rates in the presence of N protons, are then given in Eq. (S1) - Eq. (S6) (see Ref. [1] for full expressions). The relaxation rates of $\hat{S}_{0,\beta/\alpha}$ will be:

$$-R\left[\hat{S}_{0,\beta/\alpha}\right] \simeq R_{\text{local}} + R_{\text{Redfield,HFC}}$$

$$\simeq R_{\text{local}} + N \left[ \begin{array}{l} \dfrac{\Delta^2_{\Delta\text{HF}}}{180} J\left(J_{\text{ex}} \mp \omega_E \pm \omega_N\right) + \\[4pt] \dfrac{\Delta^2_{\Delta\text{HF}}}{30} J\left(J_{\text{ex}} \pm \omega_E \pm \omega_N\right) + \\[4pt] + \dfrac{\Delta^2_{\Delta\text{HF}}}{120} J\left(J_{\text{ex}} - \omega_E\right) + \dfrac{\Delta^2_{\Delta\text{HF}}}{120} J\left(J_{\text{ex}} + \omega_E\right) + \\[4pt] + \dfrac{\left[\Delta^2_{\Delta\text{HF}} + 4\aleph_{\Delta G,\Delta G \mp \Delta\text{HF}}\right]}{120} J\left(J_{\text{ex}} - \omega_E\right) + \\[4pt] + \dfrac{\left[\Delta^2_{\Delta\text{HF}} + 4\aleph_{\Delta G,\Delta G \mp \Delta\text{HF}}\right]}{120} J\left(J_{\text{ex}} + \omega_E\right) \end{array} \right] + \ldots \quad (S1)$$

Relaxation rates of $\hat{T}_{+1,\beta/\alpha}$ will be:

$$-R\left[\hat{T}_{+,\beta}\right] \simeq R_{\text{local}} + R_{\text{Redfield,HFC}}$$

$$\simeq R_{\text{local}} + N \left[ \begin{array}{l} \dfrac{\Delta^2_{\Sigma\text{HF}}}{60} J(\omega_N) + \dfrac{\Delta^2_{\Delta\text{HF}}}{180} J\left(J_{\text{ex}} + \omega_E - \omega_N\right) + \\[4pt] + \dfrac{\Delta^2_{\Delta\text{HF}}}{120} J\left(J_{\text{ex}} + \omega_E\right) + \dfrac{\left[\Delta^2_{\Delta\text{HF}} + 4\aleph_{\Delta G,\Delta G - \Delta\text{HF}}\right]}{120} J\left(J_{\text{ex}} + \omega_E\right) \end{array} \right] + \ldots \quad (S2)$$

and

$$-R\left[\hat{T}_{+,\alpha}\right] \simeq R_{\text{local}} + R_{\text{Redfield,HFC}}$$

$$\simeq R_{\text{local}} + N \left[ \begin{array}{l} \dfrac{\Delta^2_{\Sigma\text{HF}}}{60} J(\omega_N) + \dfrac{\Delta^2_{\Delta\text{HF}}}{30} J\left(J_{\text{ex}} + \omega_E + \omega_N\right) + \\[4pt] + \dfrac{\Delta^2_{\Delta\text{HF}}}{120} J\left(J_{\text{ex}} + \omega_E\right) + \dfrac{\left[\Delta^2_{\Delta\text{HF}} + 4\aleph_{\Delta G,\Delta G + \Delta\text{HF}}\right]}{120} J\left(J_{\text{ex}} + \omega_E\right) \end{array} \right] + \ldots \quad (S3)$$

Relaxation rates of $\hat{T}_{-,\beta/\alpha}$ will be:

$$-R\left[\hat{T}_{-1,\beta}\right] \simeq R_{\text{local}} + R_{\text{Redfield,HFC}}$$

$$\simeq R_{\text{local}} + N \left[ \begin{array}{l} \dfrac{\Delta^2_{\Sigma\text{HF}}}{60} J(\omega_N) + \dfrac{\Delta^2_{\Delta\text{HF}}}{30} J\left(J_{\text{ex}} - \omega_E - \omega_N\right) + \\[4pt] + \dfrac{\Delta^2_{\Delta\text{HF}}}{120} J\left(J_{\text{ex}} - \omega_E\right) + \dfrac{\left[\Delta^2_{\Delta\text{HF}} + 4\aleph_{\Delta G,\Delta G - \Delta\text{HF}}\right]}{120} J\left(J_{\text{ex}} - \omega_E\right) \end{array} \right] + \ldots \quad (S4)$$

and

$$-R\left[\hat{T}_{-1,\alpha}\right] \simeq R_{\text{local}} + R_{\text{Redfield,HFC}}$$

$$\simeq R_{\text{local}} + N\left[\begin{array}{l}\dfrac{\Delta_{\Sigma\text{HF}}^2}{60}J(\omega_N) + \dfrac{\Delta_{\Delta\text{HF}}^2}{180}J(J_{ex}-\omega_E+\omega_N)+\\ +\dfrac{\Delta_{\Delta\text{HF}}^2}{120}J(J_{ex}-\omega_E) + \dfrac{\left[\Delta_{\Delta\text{HF}}^2 + 4\aleph_{\Delta G,\Delta G+\Delta\text{HF}}\right]}{120}J(J_{ex}-\omega_E)\end{array}\right] + \ldots \quad (S5)$$

And the relaxation rates of $\hat{T}_{0,\beta/\alpha}$ correspond to:

$$-R\left[\hat{T}_{0,\beta/\alpha}\right] \simeq \dfrac{6\Delta_{EE}^2}{90}J(\omega_E) + \dfrac{6\aleph_{\Sigma G,\Sigma G\mp N(\Sigma\text{HF})}}{90}J(\omega_E) + \ldots \quad (S6)$$

where $R_{\text{local}}$ represents the relaxation term arising from local vibrational modes[57, 58, 67]; $R_{\text{Redfield,HFC}}$ represents the Redfield relaxation term arising from the dipolar hyperfine interactions between the electrons and a number of intra-radical protons N; the $\Delta_{\Sigma\text{HF}}^2$ and $\Delta_{\Delta\text{HF}}^2$ terms are the second rank norm squared arising from anisotropies associated to the sum and the difference between the hyperfine coupling tensors between the protons and the two electrons; $\Delta_{EE}^2$ is the second rank norm squared arising from the anisotropy present in the inter-electron dipolar coupling tensor, $\aleph_{\Sigma G,\Sigma G\mp N(\Sigma\text{HF})}$ is the second-rank scalar product between two 3×3 interaction tensors arising from the sum of the two $g$-tensors and the sum of the two $g$-tensors ± the $\Sigma\text{HF}$ term.

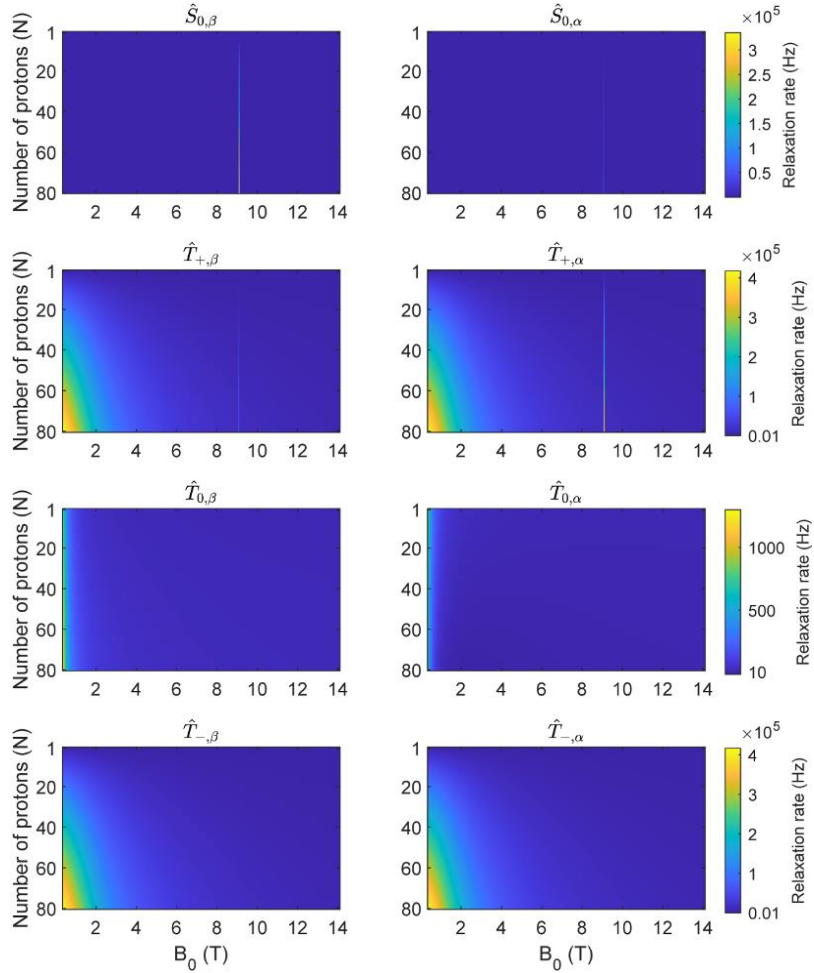

**FIG. S1:** Analytical self-relaxation rates computed in according to Eq. (S1) – (S6), with $R_{\text{local}}$= 0 Hz, as a function of the number of protons end of the magnetic field. The J-DNP condition is fulfilled at 9.1 T, the other simulation parameters are given in the Table 1 in the main text.

Fig. S1 shows the predictions of these equations for a biradical containing protons that are placed 5 Å away from one of the electrons, and 19.4 Å away from the other. These coordinates are representative of an intra-radical proton positioned in the area "C" of Fig. 1A; notice that the presence of multiple protons simply amplifies linearly the rates predicted for a single proton. At magnetic fields ≥ 3.4 T, these plots predict that the $\hat{T}_{\pm1,\beta/\alpha}$'s $T_1$(s) will range between 100s μs and 1000 μs, while the $T_1$(s) of $\hat{T}_{0,\beta/\alpha}$ range between 10s ms and 100s ms; the rates of the singlet states are zero at any field, but increase suddenly at the J-DNP condition.

Fig. S2 shows the singlet and the triplet states self-relaxation rates, if a term $R_{local}$ arising from local vibrational mode of the kind that dominate the longitudinal relaxation rates in trityl and nitroxide radicals,[57, 58, 67] is added to the scenario of Fig. S1. As can be seen from the figure below, this term is expected to affect only the relaxation rates of $\hat{T}_{\pm1,\beta/\alpha}$, since it is these states that describe the longitudinal electron relaxation. The presence of this local vibrational relaxation mode makes the $\hat{T}_{\pm1,\beta/\alpha}$'s $T_1$(s) drop to tens of μs. However, a population imbalance can still be observed at the JDNP condition between $\hat{S}_{0,\beta}$ and $\hat{S}_{0,\alpha}$, and $\hat{T}_{+1,\beta}$ and $\hat{T}_{+1,\alpha}$.

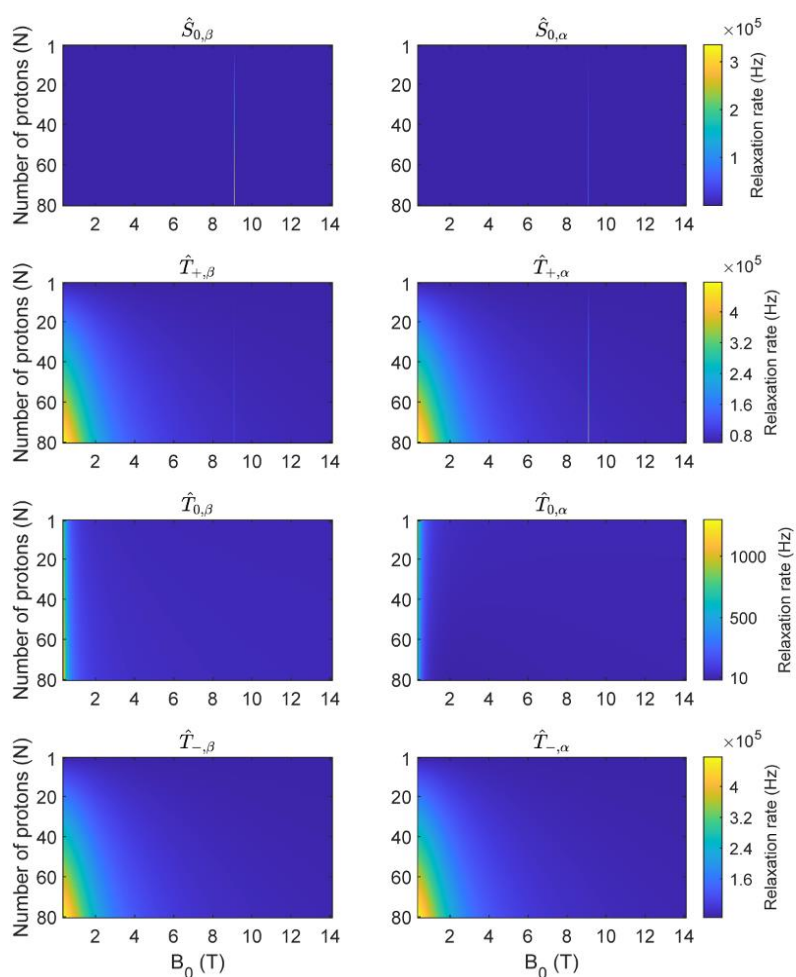

**FIG. S2:** Analytical self-relaxation rates computed in according to Eq. (S1) – (S6), with $R_{local}$ = 0.6 x $10^5$ Hz, as a function of N end of the magnetic field. The J-DNP condition is fulfilled at 9.1 T, the other simulation parameters are given in the Table 1 in the main text.

## C. Numerical singlet and triplet relaxation rates as a function of the magnetic field

As discussed in Ref. [30], triplet and singlet relaxation rates can be orders-of-magnitude smaller than longitudinal $T_1$ electron relaxation rates –which reach in excess of ≈$10^6$ Hz in biradicals at any magnetic field. The corresponding to $T_1$s of orders-of-magnitude shorter than the assumed shuttling times, the reason why MF-JDNP still enhances nuclear polarization in such case, relates to the fact that the electron's singlet and triplet relaxation rates that are involved in the process, will be order of magnitudes smaller than the apparent electronic $1/T_1$ rate, this is illustrated in FIG. S3.

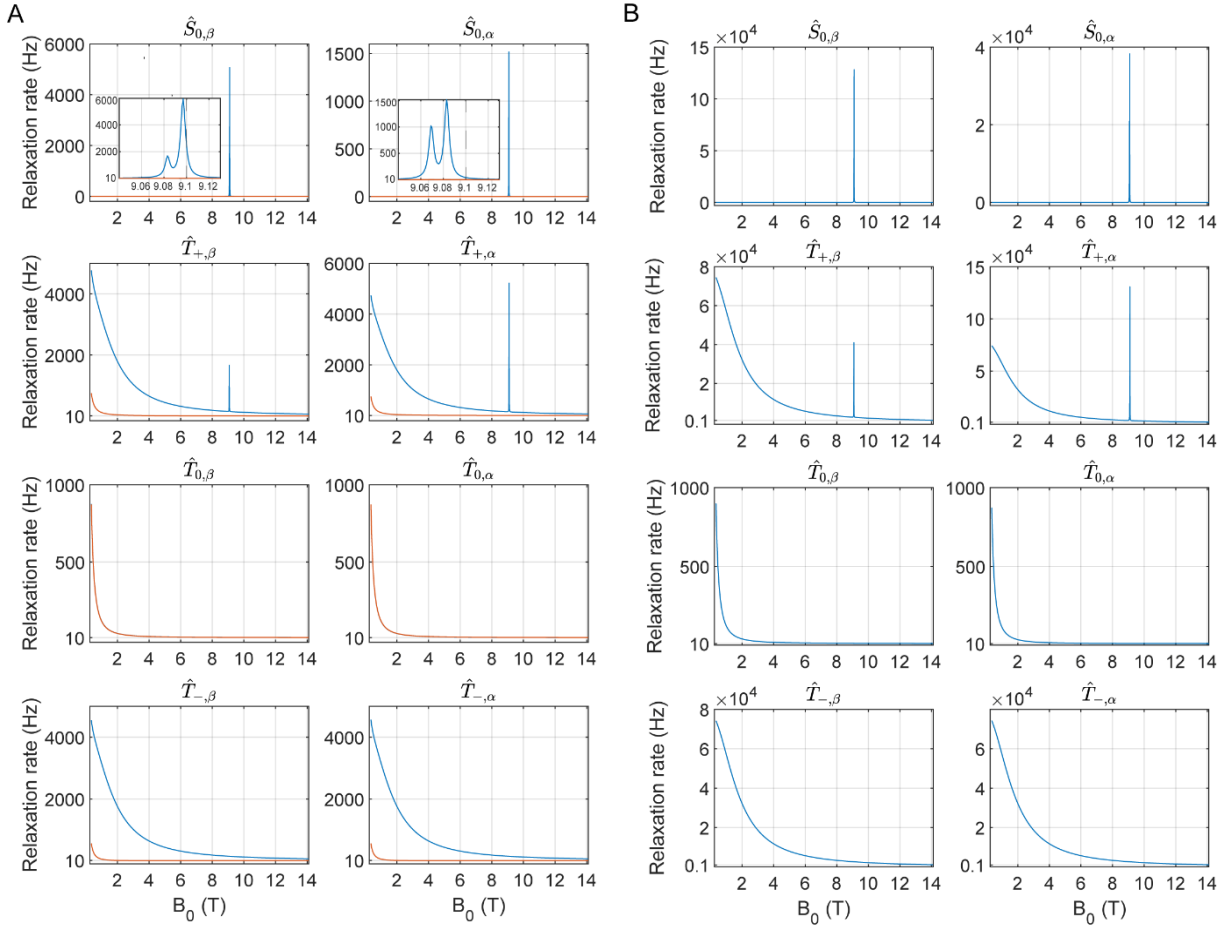

**FIG. S3:** Redfield self-relaxation rates of the α and β nuclear components of the singlet and the triplet states as a function of the magnetic field for: (A) a three-spin system with proton set in the configuration A (blue lines) and the proton set in the configuration B (red lines); (B) a three-spin system considering the coordinates of the intra-radical proton set in the configuration C. The JDNP was set at 9.1 T, indicated with a dashed grey line. The three peaks in the zoomed region correspond to the three JDNP matching conditions arising from the terms $J(J_{ex} \pm \omega_E \pm \omega_N)$, $J(J_{ex} \pm \omega_E)$ and $J(J_{ex} \pm \omega_E \mp \omega_N)$ in the analytical expression describing the self-relaxation rates are shown in [30].

The rates of the $\hat{T}_{\pm,\alpha/\beta}$ triplet states decrease with the magnetic field but have a singular increase at the JDNP condition, leading to different fates for the α and β nuclear spin states (FIG. S3). The closer is the proton to one of the electrons and the stronger is the dipolar hyperfine relaxation, notice the decrease of the relaxation rate from the proton set in the configuration C to the proton set in the configuration A. Much slower and hyperfine-independent self-relaxation rates were observed for $\hat{T}_{0,\alpha/\beta}$, while the self-relaxation rates of the singlet states remain virtually at zero at all fields in which the JDNP condition is not fulfilled. The rate at which the maximum nuclear magnetization will build up will then be determined by the relaxation rates of $\hat{S}_{0,\beta}$ and $\hat{T}_{+,\alpha}$, while its decay is determined by the rates $\hat{S}_{0,\alpha}$ and $\hat{T}_{+,\beta}$, in particular $\hat{S}_{0,\alpha}$, that has a lower relaxation rate. Shuttling rates faster than the latter self-relaxation rates at the polarization field is thus all that is needed, since as the sample is then moved to higher fields the self-relaxation rates drop further or even become zero. This helps to "freeze" the decay of the enhancement obtained at the JDNP condition, and maintain it significant at the NMR field where the measurement is performed.

### D. Effect of *g*-tensor orientation on the enhancement

In the case of axially symmetric *g*-tensors of symmetric biradicals, given a directional cosine matrix and using the active ZYZ Euler rotation convention $\mathbf{VDV}^{-1}$ [68], where $\mathbf{V}$ is the rotational matrices and $\mathbf{D}$ is the interaction tensor, it follows that rotations about the Y β-angle (on the linker connecting the two monomeric units in a symmetric biradical) can lead to a variation. Due to the axial symmetry, no variations can be observed upon the rotation about the X γ-angle and Z α-angle. A difference between the two anisotropic *g*-tensors, will lead to the $\aleph_{\Delta G, \Delta G \pm \Delta HF}$ terms to overtake the $\Delta_{HFC}^2$ in Eqs. (S1) – (S5), that will lead to an imbalance between the α and β components of the singlet and the triplet states, robbing efficiency from JDNP. However, assuming an aromatic or multiple-bond linker between radical units that can bend with a limited |β| ≤9° excursion, Figure S4 shows the impact that this breaking of the collinearity will have on the MF-JDNP enhancement. As can be appreciated by comparing this Figure with the data shown in Figure 4, the effect of such bending distortion would be minimal.

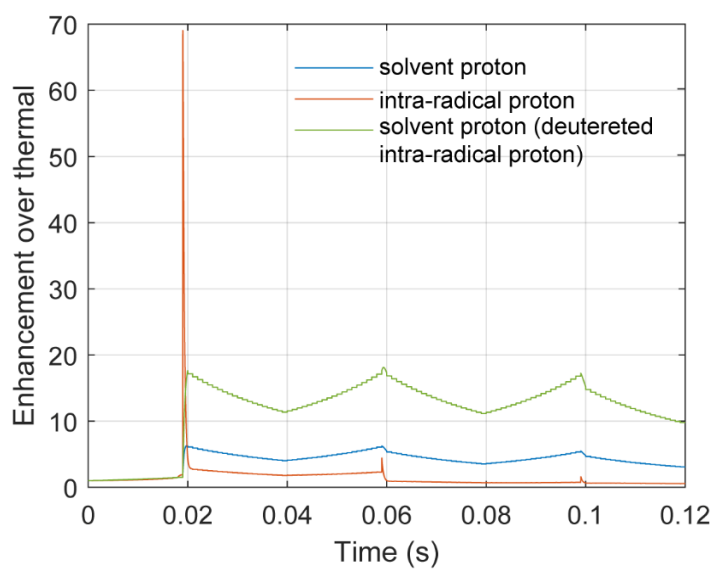

**Figure S4:** Expectations of MF-JDNP experiments performed according to the scheme in Fig. 2 in the main text, with three high-low-high $B_0$ shuttling repetitions. All plots show time/magnetic field evolution of the nuclear enhancement over the thermal equilibrium value. Predictions were performed using a four-spin system including a fixed intra-radical proton and a diffusion "solvent", as described in FIG. 4, but using non-collinear *g*-tensors, that were maintained axially symmetric along the main molecular axis (corresponding the linker connecting the two mono-radical units). The first *g*-tensor's eigenvalues were set to [2.0032 2.0032 2.0026] and the respective ZYZ Euler angles were set to [π/4 π/20 π/2] rad, while second *g*-tensor's eigenvalues were set to [2.0032 2.0032 2.0026] and the respective ZYZ Euler angles were set to [0 0 0] rad.